\documentclass[preliminary,copyright,creativecommons]{eptcs}
\providecommand{\event}{GandALF 2017} 
\usepackage{breakurl}             
\usepackage{underscore}           
\usepackage[svgnames]{xcolor}
\usepackage{amssymb}

\usepackage[utf8]{inputenc}

\DeclareUnicodeCharacter{00AC}{\lnot}
\DeclareUnicodeCharacter{00B1}{\pm}
\DeclareUnicodeCharacter{00D7}{\times}

\usepackage{isabelle}
\usepackage{isabellesym}
\newcommand{\DefineSnippet}[2]{%
   \expandafter\newcommand\csname snippet--#1\endcsname{%
\begingroup\setlength{\leftmargini}{0em}
     \begin{quote}
\fcolorbox[rgb]{0.85,0.85,0.86}[rgb]{0.85,0.85,0.86}{\parbox{0.985\textwidth}{\noindent
     \begin{isabelle} 
     #2
     \end{isabelle}}}
     \end{quote}\endgroup}}
\newcommand{\Snippet}[1]{\csname snippet--#1\endcsname}
\newcommand{\DefineSnippetInline}[2]{%
   \expandafter\newcommand\csname snippet--#1\endcsname{%
\begingroup\setlength{\leftmargini}{0em}
\fcolorbox[rgb]{0.85,0.85,0.86}[rgb]{0.85,0.85,0.86}{\noindent
\isa{#2}}
     \endgroup}}
 
\DefineSnippet{uconelessP}{
\isacommand{definition}\isamarkupfalse%
\isanewline
\ \ ucone\ {\isacharcolon}{\isacharcolon}\ {\isachardoublequoteopen}{\isacharparenleft}{\isacharprime}a\ {\isasymRightarrow}\ {\isacharprime}a\ {\isasymRightarrow}\ bool{\isacharparenright}\ {\isasymRightarrow}\ {\isacharprime}a\ {\isasymRightarrow}\ {\isacharparenleft}{\isacharprime}a\ set{\isacharparenright}{\isachardoublequoteclose}\isanewline
\ \ \isakeyword{where}\ {\isachardoublequoteopen}ucone\ les\ x\ {\isacharequal}\ {\isacharbraceleft}z{\isachardot}\ les\ x\ z{\isacharbraceright}{\isachardoublequoteclose}\ \isanewline
\isanewline
\isacommand{definition}\isamarkupfalse%
\isanewline
\ \ uCone\ {\isacharcolon}{\isacharcolon}\ {\isachardoublequoteopen}{\isacharparenleft}{\isacharprime}a\ {\isasymRightarrow}\ {\isacharprime}a\ {\isasymRightarrow}\ bool{\isacharparenright}\ {\isasymRightarrow}\ {\isacharparenleft}{\isacharprime}a\ set{\isacharparenright}\ {\isasymRightarrow}\ {\isacharparenleft}{\isacharprime}a\ set{\isacharparenright}{\isachardoublequoteclose}\isanewline
\ \ \isakeyword{where}\ {\isachardoublequoteopen}uCone\ les\ Y\ {\isacharequal}\ {\isasymUnion}\ {\isacharparenleft}{\isacharparenleft}ucone\ les{\isacharparenright}\ {\isacharbackquote}\ Y{\isacharparenright}{\isachardoublequoteclose}\ \isanewline
\isanewline
\isacommand{definition}\isamarkupfalse%
\isanewline
\ \ lessP\ {\isacharcolon}{\isacharcolon}\ {\isachardoublequoteopen}{\isacharparenleft}{\isacharprime}a\ {\isasymRightarrow}\ {\isacharprime}a\ {\isasymRightarrow}\ bool{\isacharparenright}\ {\isasymRightarrow}\ {\isacharparenleft}{\isacharprime}a\ set{\isacharparenright}\ {\isasymRightarrow}\ {\isacharparenleft}{\isacharprime}a\ set{\isacharparenright}\ {\isasymRightarrow}\ bool{\isachardoublequoteclose}\isanewline
\ \ \isakeyword{where}\ {\isachardoublequoteopen}lessP\ les\ A\ B\ {\isacharequal}\ {\isacharparenleft}finite\ A\ {\isasymand}\ \isanewline
\ \ \ \ \ {\isacharparenleft}{\isasymexists}A{\isacharprime}{\isasymsubseteq}A{\isacharminus}B{\isachardot}\ A{\isacharprime}{\isasymnoteq}{\isacharbraceleft}{\isacharbraceright}\ {\isasymand}\ A{\isacharminus}{\isacharparenleft}A{\isacharprime}{\isasymunion}{\isacharparenleft}uCone\ les\ A{\isacharprime}{\isacharparenright}{\isacharparenright}\ {\isacharequal}\ B{\isacharminus}{\isacharparenleft}A{\isacharprime}{\isasymunion}{\isacharparenleft}uCone\ les\ A{\isacharprime}{\isacharparenright}{\isacharparenright}{\isacharparenright}{\isacharparenright}{\isachardoublequoteclose}%
}
\DefineSnippet{liftirreflexive}{
\isacommand{lemma}\isamarkupfalse%
\ lift{\isacharunderscore}irreflexive\ {\isacharcolon}\ \isanewline
\ \ \isakeyword{shows}\ {\isachardoublequoteopen}{\isasymnot}{\isacharparenleft}lessP\ les\ A\ A{\isacharparenright}{\isachardoublequoteclose}%
}
\DefineSnippet{lifttransitive}{
\isacommand{lemma}\isamarkupfalse%
\ lift{\isacharunderscore}transitive\ {\isacharcolon}\ \ \ \isanewline
\ \ \isakeyword{assumes}\ les{\isacharunderscore}irr\ {\isacharcolon}\ {\isachardoublequoteopen}{\isasymAnd}x{\isachardot}\ {\isasymnot}\ {\isacharparenleft}les\ x\ x{\isacharparenright}{\isachardoublequoteclose}\isanewline
\ \ \ \ \ \ \isakeyword{and}\ les{\isacharunderscore}trans\ {\isacharcolon}\ {\isachardoublequoteopen}{\isasymAnd}x\ y\ z{\isachardot}\ {\isacharparenleft}les\ x\ y{\isacharparenright}\ {\isasymLongrightarrow}\ {\isacharparenleft}les\ y\ z{\isacharparenright}\ {\isasymLongrightarrow}\ {\isacharparenleft}les\ x\ z{\isacharparenright}{\isachardoublequoteclose}\isanewline
\ \ \ \ \ \ \isakeyword{and}\ AB\ {\isacharcolon}\ {\isachardoublequoteopen}lessP\ les\ A\ B{\isachardoublequoteclose}\isanewline
\ \ \ \ \ \ \isakeyword{and}\ BC\ {\isacharcolon}\ {\isachardoublequoteopen}lessP\ les\ B\ C{\isachardoublequoteclose}\ \ \ \ \ \isanewline
\ \ \isakeyword{shows}\ {\isachardoublequoteopen}lessP\ les\ A\ C{\isachardoublequoteclose}%
}
\DefineSnippet{replaceWithPreferred}{
\isacommand{definition}\isamarkupfalse%
\ \isanewline
\ \ replaceWithPreferred\ {\isacharcolon}{\isacharcolon}\ {\isachardoublequoteopen}{\isacharparenleft}{\isacharprime}a\ {\isasymRightarrow}\ {\isacharprime}a\ {\isasymRightarrow}\ bool{\isacharparenright}\ {\isasymRightarrow}\ {\isacharprime}a\ {\isasymRightarrow}\ {\isacharprime}a\ set\ {\isasymRightarrow}\ {\isacharprime}a\ set\ {\isasymRightarrow}\ {\isacharprime}a\ set{\isachardoublequoteclose}\isanewline
\ \ \isakeyword{where}\ {\isachardoublequoteopen}replaceWithPreferred\ les\ a\ A\ U\ {\isacharequal}\ {\isacharparenleft}A\ {\isasymunion}\ {\isacharbraceleft}a{\isacharprime}{\isasymin}U{\isachardot}\ les\ a\ a{\isacharprime}{\isacharbraceright}{\isacharparenright}{\isacharminus}{\isacharbraceleft}a{\isacharbraceright}{\isachardoublequoteclose}%
}
\DefineSnippet{replaceIsPreferred}{
\isacommand{lemma}\isamarkupfalse%
\ replaceIsPreferred\ {\isacharcolon}\ \isanewline
\ \ \isakeyword{assumes}\ a{\isacharunderscore}A\ {\isacharcolon}\ {\isachardoublequoteopen}a{\isasymin}A{\isachardoublequoteclose}\isanewline
\ \ \isakeyword{and}\ finiteA\ {\isacharcolon}\ {\isachardoublequoteopen}finite\ A{\isachardoublequoteclose}\isanewline
\ \ \isakeyword{shows}\ {\isachardoublequoteopen}lessP\ les\ A\ {\isacharparenleft}replaceWithPreferred\ les\ a\ A\ U{\isacharparenright}{\isachardoublequoteclose}%
}
\DefineSnippet{gameform}{
\isacommand{type{\isacharunderscore}synonym}\isamarkupfalse%
\ {\isacharparenleft}{\isacharprime}O{\isacharcomma}{\isacharprime}S{\isadigit{1}}{\isacharcomma}{\isacharprime}S{\isadigit{2}}{\isacharparenright}\ game{\isacharunderscore}form\ {\isacharequal}\ {\isachardoublequoteopen}{\isacharparenleft}{\isacharprime}S{\isadigit{1}}\ {\isacharasterisk}\ {\isacharprime}S{\isadigit{2}}{\isacharparenright}\ {\isasymRightarrow}\ {\isacharprime}O{\isachardoublequoteclose}%
}
\DefineSnippet{game}{
\isacommand{type{\isacharunderscore}synonym}\isamarkupfalse%
\ {\isacharparenleft}{\isacharprime}O{\isacharcomma}{\isacharprime}S{\isadigit{1}}{\isacharcomma}{\isacharprime}S{\isadigit{2}}{\isacharparenright}\ game\ {\isacharequal}\ \isanewline
\ \ \ {\isachardoublequoteopen}{\isacharparenleft}{\isacharprime}O\ {\isasymRightarrow}\ {\isacharprime}O\ {\isasymRightarrow}\ bool{\isacharparenright}\ {\isacharasterisk}\ {\isacharparenleft}{\isacharprime}O\ {\isasymRightarrow}\ {\isacharprime}O\ {\isasymRightarrow}\ bool{\isacharparenright}\ {\isacharasterisk}\ {\isacharparenleft}{\isacharparenleft}{\isacharprime}O{\isacharcomma}{\isacharprime}S{\isadigit{1}}{\isacharcomma}{\isacharprime}S{\isadigit{2}}{\isacharparenright}\ game{\isacharunderscore}form{\isacharparenright}{\isachardoublequoteclose}%
}
\DefineSnippet{isNash}{
\isacommand{definition}\isamarkupfalse%
\isanewline
\ \ isNash\ {\isacharcolon}{\isacharcolon}\ {\isachardoublequoteopen}{\isacharparenleft}{\isacharparenleft}{\isacharprime}O{\isacharcomma}{\isacharprime}S{\isadigit{1}}{\isacharcomma}{\isacharprime}S{\isadigit{2}}{\isacharparenright}\ game{\isacharparenright}\ {\isasymRightarrow}\ {\isacharprime}S{\isadigit{1}}\ {\isasymRightarrow}\ {\isacharprime}S{\isadigit{2}}\ {\isasymRightarrow}\ bool{\isachardoublequoteclose}\isanewline
\ \ \isakeyword{where}\ {\isachardoublequoteopen}isNash\ g\ s{\isadigit{1}}\ s{\isadigit{2}}\ {\isacharequal}\ \isanewline
\ \ \ \ \ {\isacharparenleft}{\isacharparenleft}{\isasymforall}s{\isadigit{1}}{\isacharprime}{\isachardot}\ {\isasymnot}{\isacharparenleft}pref{\isadigit{1}}\ g{\isacharparenright}\ {\isacharparenleft}{\isacharparenleft}form\ g{\isacharparenright}\ {\isacharparenleft}s{\isadigit{1}}{\isacharcomma}s{\isadigit{2}}{\isacharparenright}{\isacharparenright}\ {\isacharparenleft}{\isacharparenleft}form\ g{\isacharparenright}\ {\isacharparenleft}s{\isadigit{1}}{\isacharprime}{\isacharcomma}s{\isadigit{2}}{\isacharparenright}{\isacharparenright}{\isacharparenright}\ {\isasymand}\isanewline
\ \ \ \ \ \ {\isacharparenleft}{\isasymforall}s{\isadigit{2}}{\isacharprime}{\isachardot}\ {\isasymnot}{\isacharparenleft}pref{\isadigit{2}}\ g{\isacharparenright}\ {\isacharparenleft}{\isacharparenleft}form\ g{\isacharparenright}\ {\isacharparenleft}s{\isadigit{1}}{\isacharcomma}s{\isadigit{2}}{\isacharparenright}{\isacharparenright}\ {\isacharparenleft}{\isacharparenleft}form\ g{\isacharparenright}\ {\isacharparenleft}s{\isadigit{1}}{\isacharcomma}s{\isadigit{2}}{\isacharprime}{\isacharparenright}{\isacharparenright}{\isacharparenright}{\isacharparenright}{\isachardoublequoteclose}%
}
\DefineSnippet{determined}{
\isacommand{definition}\isamarkupfalse%
\ \isanewline
\ \ determined\ {\isacharcolon}{\isacharcolon}\ {\isachardoublequoteopen}{\isacharparenleft}{\isacharparenleft}bool{\isacharcomma}{\isacharprime}S{\isadigit{1}}{\isacharcomma}{\isacharprime}S{\isadigit{2}}{\isacharparenright}\ game{\isacharparenright}\ {\isasymRightarrow}\ {\isacharparenleft}{\isacharprime}S{\isadigit{1}}\ set{\isacharparenright}\ {\isasymRightarrow}\ {\isacharparenleft}{\isacharprime}S{\isadigit{2}}\ set{\isacharparenright}\ {\isasymRightarrow}\ bool{\isachardoublequoteclose}\isanewline
\ \ \isakeyword{where}\ {\isachardoublequoteopen}determined\ g\ R{\isadigit{1}}\ R{\isadigit{2}}\ {\isacharequal}\ {\isacharparenleft}{\isacharparenleft}{\isasymexists}s{\isadigit{1}}{\isasymin}R{\isadigit{1}}{\isachardot}\ {\isasymforall}s{\isadigit{2}}{\isachardot}\ {\isacharparenleft}form\ g{\isacharparenright}\ {\isacharparenleft}s{\isadigit{1}}{\isacharcomma}s{\isadigit{2}}{\isacharparenright}\ {\isacharequal}\ True{\isacharparenright}\ {\isasymor}\isanewline
\ \ \ \ \ \ \ \ \ \ \ \ \ \ \ \ \ \ \ \ \ \ \ \ \ \ \ \ \ \ \ {\isacharparenleft}{\isasymexists}s{\isadigit{2}}{\isasymin}R{\isadigit{2}}{\isachardot}\ {\isasymforall}s{\isadigit{1}}{\isachardot}\ {\isacharparenleft}form\ g{\isacharparenright}\ {\isacharparenleft}s{\isadigit{1}}{\isacharcomma}s{\isadigit{2}}{\isacharparenright}\ {\isacharequal}\ False{\isacharparenright}{\isacharparenright}{\isachardoublequoteclose}%
}
\DefineSnippet{derivedWLGame}{
\isacommand{definition}\isamarkupfalse%
\isanewline
\ \ derivedWLGame\ {\isacharcolon}{\isacharcolon}\ {\isachardoublequoteopen}{\isacharparenleft}{\isacharparenleft}{\isacharprime}O{\isacharcomma}{\isacharprime}S{\isadigit{1}}{\isacharcomma}{\isacharprime}S{\isadigit{2}}{\isacharparenright}\ game{\isacharunderscore}form{\isacharparenright}\ {\isasymRightarrow}\ {\isacharparenleft}{\isacharprime}O\ set{\isacharparenright}\ {\isasymRightarrow}\ {\isacharparenleft}{\isacharparenleft}bool{\isacharcomma}{\isacharprime}S{\isadigit{1}}{\isacharcomma}{\isacharprime}S{\isadigit{2}}{\isacharparenright}\ game{\isacharparenright}{\isachardoublequoteclose}\isanewline
\ \ \isakeyword{where}\ {\isachardoublequoteopen}derivedWLGame\ gf\ Ou\ {\isacharequal}\ \isanewline
\ \ \ \ \ {\isacharparenleft}{\isacharparenleft}{\isasymlambda}\ ou\ p{\isachardot}\ p{\isasymand}{\isasymnot}ou{\isacharparenright}{\isacharcomma}\ {\isacharparenleft}{\isasymlambda}\ ou\ p{\isachardot}\ ou{\isasymand}{\isasymnot}p{\isacharparenright}{\isacharcomma}\ {\isacharparenleft}{\isasymlambda}ou{\isachardot}\ ou{\isasymin}Ou{\isacharparenright}{\isasymcirc}gf{\isacharparenright}{\isachardoublequoteclose}%
}
\DefineSnippet{someonewins}{
\isacommand{lemma}\isamarkupfalse%
\ someone{\isacharunderscore}wins\ {\isacharcolon}\ \isanewline
\ \ \isakeyword{assumes}\ isNashWL\ {\isacharcolon}\ {\isachardoublequoteopen}isNash\ {\isacharparenleft}{\isacharparenleft}derivedWLGame\ gf\ Ou{\isacharparenright}{\isacharcolon}{\isacharcolon}{\isacharparenleft}{\isacharparenleft}bool{\isacharcomma}{\isacharprime}S{\isadigit{1}}{\isacharcomma}{\isacharprime}S{\isadigit{2}}{\isacharparenright}\ game{\isacharparenright}{\isacharparenright}\ s{\isadigit{1}}\ s{\isadigit{2}}{\isachardoublequoteclose}\ \isanewline
\ \ \ \ \ {\isacharparenleft}\isakeyword{is}\ {\isachardoublequoteopen}isNash\ {\isacharquery}wlG\ s{\isadigit{1}}\ s{\isadigit{2}}{\isachardoublequoteclose}{\isacharparenright}\isanewline
\ \ \isakeyword{shows}\ {\isachardoublequoteopen}{\isacharparenleft}{\isasymforall}s{\isadigit{2}}{\isacharprime}{\isachardot}\ {\isacharparenleft}form\ {\isacharquery}wlG{\isacharparenright}\ {\isacharparenleft}s{\isadigit{1}}{\isacharcomma}s{\isadigit{2}}{\isacharprime}{\isacharparenright}\ {\isacharequal}\ True{\isacharparenright}\ {\isasymor}\ {\isacharparenleft}{\isasymforall}s{\isadigit{1}}{\isacharprime}{\isachardot}\ {\isacharparenleft}form\ {\isacharquery}wlG{\isacharparenright}\ {\isacharparenleft}s{\isadigit{1}}{\isacharprime}{\isacharcomma}s{\isadigit{2}}{\isacharparenright}\ {\isacharequal}\ False{\isacharparenright}{\isachardoublequoteclose}%
}
\DefineSnippet{determinedForm}{
\isacommand{definition}\isamarkupfalse%
\ \isanewline
\ \ determinedForm\ {\isacharcolon}{\isacharcolon}\ {\isachardoublequoteopen}{\isacharparenleft}{\isacharparenleft}{\isacharprime}O{\isacharcomma}{\isacharprime}S{\isadigit{1}}{\isacharcomma}{\isacharprime}S{\isadigit{2}}{\isacharparenright}\ game{\isacharunderscore}form{\isacharparenright}\ {\isasymRightarrow}\ {\isacharparenleft}{\isacharprime}S{\isadigit{1}}\ set{\isacharparenright}\ {\isasymRightarrow}\ {\isacharparenleft}{\isacharprime}S{\isadigit{2}}\ set{\isacharparenright}\ {\isasymRightarrow}\ bool{\isachardoublequoteclose}\isanewline
\ \ \isakeyword{where}\ {\isachardoublequoteopen}determinedForm\ gf\ R{\isadigit{1}}\ R{\isadigit{2}}\ {\isacharequal}\ {\isacharparenleft}{\isasymforall}\ Ou{\isachardot}\ determined\ {\isacharparenleft}derivedWLGame\ gf\ Ou{\isacharparenright}\ R{\isadigit{1}}\ R{\isadigit{2}}{\isacharparenright}{\isachardoublequoteclose}%
}
\DefineSnippet{enforceSet}{
\isacommand{definition}\isamarkupfalse%
\isanewline
\ \ enforceSet{\isadigit{1}}\ {\isacharcolon}{\isacharcolon}\ \ {\isachardoublequoteopen}{\isacharparenleft}{\isacharparenleft}{\isacharprime}O{\isacharcomma}{\isacharprime}S{\isadigit{1}}{\isacharcomma}{\isacharprime}S{\isadigit{2}}{\isacharparenright}\ game{\isacharunderscore}form{\isacharparenright}\ {\isasymRightarrow}\ {\isacharprime}S{\isadigit{1}}\ {\isasymRightarrow}\ {\isacharparenleft}{\isacharprime}O\ set{\isacharparenright}{\isachardoublequoteclose}\isanewline
\ \ \isakeyword{where}\ {\isachardoublequoteopen}enforceSet{\isadigit{1}}\ f\ s{\isadigit{1}}\ {\isacharequal}\ {\isacharbraceleft}ou{\isachardot}\ {\isasymexists}s{\isadigit{2}}{\isachardot}\ f{\isacharparenleft}s{\isadigit{1}}{\isacharcomma}s{\isadigit{2}}{\isacharparenright}\ {\isacharequal}\ ou{\isacharbraceright}{\isachardoublequoteclose}%
}
\DefineSnippet{enforceSets}{
\isacommand{definition}\isamarkupfalse%
\isanewline
\ \ enforceSets{\isadigit{1}}\ {\isacharcolon}{\isacharcolon}\ \ {\isachardoublequoteopen}{\isacharparenleft}{\isacharparenleft}{\isacharprime}O{\isacharcomma}{\isacharprime}S{\isadigit{1}}{\isacharcomma}{\isacharprime}S{\isadigit{2}}{\isacharparenright}\ game{\isacharunderscore}form{\isacharparenright}\ {\isasymRightarrow}\ {\isacharparenleft}{\isacharprime}O\ set\ set{\isacharparenright}{\isachardoublequoteclose}\isanewline
\ \ \isakeyword{where}\ {\isachardoublequoteopen}enforceSets{\isadigit{1}}\ f\ {\isacharequal}\ range\ {\isacharparenleft}enforceSet{\isadigit{1}}\ f{\isacharparenright}{\isachardoublequoteclose}%
}
\DefineSnippet{determinedFormViaImplEnforceOutc}{
\isacommand{lemma}\isamarkupfalse%
\ determined{\isacharunderscore}form{\isacharunderscore}via{\isacharunderscore}impl{\isacharunderscore}enforce{\isacharunderscore}outc\ {\isacharcolon}\ \isanewline
\ \ \isakeyword{assumes}\ det\ {\isacharcolon}\ {\isachardoublequoteopen}determinedForm\ gf\ R{\isadigit{1}}\ R{\isadigit{2}}{\isachardoublequoteclose}\isanewline
\ \ \isakeyword{shows}\ {\isachardoublequoteopen}{\isacharparenleft}{\isasymforall}\ Ou{\isachardot}\ {\isacharparenleft}{\isasymexists}s{\isadigit{1}}{\isasymin}R{\isadigit{1}}{\isachardot}\ enforceSet{\isadigit{1}}\ gf\ s{\isadigit{1}}\ {\isasymsubseteq}\ Ou{\isacharparenright}{\isasymor}\isanewline
\ \ \ \ \ \ \ \ \ \ \ \ \ \ \ \ {\isacharparenleft}{\isasymexists}s{\isadigit{2}}{\isasymin}R{\isadigit{2}}{\isachardot}\ enforceSet{\isadigit{2}}\ gf\ s{\isadigit{2}}\ {\isasymsubseteq}\ {\isacharparenleft}range\ gf{\isacharparenright}\ {\isacharminus}\ Ou{\isacharparenright}{\isacharparenright}{\isachardoublequoteclose}%
}
\DefineSnippet{equilibriumtransferfinite}{
\isacommand{theorem}\isamarkupfalse%
\ equilibrium{\isacharunderscore}transfer{\isacharunderscore}finite\ {\isacharcolon}\isanewline
\ \ \isakeyword{assumes}\ finiteO\ {\isacharcolon}\ {\isachardoublequoteopen}finite\ {\isacharparenleft}range\ {\isacharparenleft}form\ g{\isacharparenright}{\isacharparenright}{\isachardoublequoteclose}\isanewline
\ \ \ \ \ \ \isakeyword{and}\ trans{\isadigit{1}}\ {\isacharcolon}\ {\isachardoublequoteopen}{\isasymAnd}a\ b\ c{\isachardot}\ {\isacharparenleft}pref{\isadigit{1}}\ g{\isacharparenright}\ a\ b\ {\isasymLongrightarrow}\ {\isacharparenleft}pref{\isadigit{1}}\ g{\isacharparenright}\ b\ c\ {\isasymLongrightarrow}\ {\isacharparenleft}pref{\isadigit{1}}\ g{\isacharparenright}\ a\ c{\isachardoublequoteclose}\isanewline
\ \ \ \ \ \ \isakeyword{and}\ irref{\isadigit{1}}\ {\isacharcolon}\ {\isachardoublequoteopen}{\isasymAnd}a{\isachardot}\ {\isasymnot}{\isacharparenleft}pref{\isadigit{1}}\ g{\isacharparenright}\ a\ a{\isachardoublequoteclose}\ \isanewline
\ \ \ \ \ \ \isakeyword{and}\ trans{\isadigit{2}}\ {\isacharcolon}\ {\isachardoublequoteopen}{\isasymAnd}a\ b\ c{\isachardot}\ {\isacharparenleft}pref{\isadigit{2}}\ g{\isacharparenright}\ a\ b\ {\isasymLongrightarrow}\ {\isacharparenleft}pref{\isadigit{2}}\ g{\isacharparenright}\ b\ c\ {\isasymLongrightarrow}\ {\isacharparenleft}pref{\isadigit{2}}\ g{\isacharparenright}\ a\ c{\isachardoublequoteclose}\isanewline
\ \ \ \ \ \ \isakeyword{and}\ irref{\isadigit{2}}\ {\isacharcolon}\ {\isachardoublequoteopen}{\isasymAnd}a{\isachardot}\ {\isasymnot}{\isacharparenleft}pref{\isadigit{2}}\ g{\isacharparenright}\ a\ a{\isachardoublequoteclose}\ \isanewline
\ \ \ \ \ \ \isakeyword{and}\ det\ {\isacharcolon}\ {\isachardoublequoteopen}determinedForm\ {\isacharparenleft}form\ g{\isacharparenright}\ R{\isadigit{1}}\ R{\isadigit{2}}{\isachardoublequoteclose}\ \isanewline
\ \ \isakeyword{shows}\ {\isachardoublequoteopen}{\isasymexists}s{\isadigit{1}}{\isasymin}R{\isadigit{1}}{\isachardot}\ {\isasymexists}s{\isadigit{2}}{\isasymin}R{\isadigit{2}}{\isachardot}\ isNash\ g\ s{\isadigit{1}}\ s{\isadigit{2}}{\isachardoublequoteclose}%
}

\usepackage[standard,thmmarks]{ntheorem}

\theoremstyle{plain}
\theorembodyfont{\upshape}
\renewtheorem{theorem}{Theorem}
\renewtheorem{corollary}[theorem]{Corollary}
\renewtheorem{lemma}[theorem]{Lemma}
\theoremsymbol{\ensuremath{\lhd}}
\renewtheorem{remark}[theorem]{Remark}

\theoremstyle{definition}
\theoremsymbol{}
\renewtheorem{definition}[theorem]{Definition}

\theoremheaderfont{\sc}\theorembodyfont{\upshape}
\theoremstyle{nonumberplain}
\theoremsymbol{\ensuremath{\Box}}
\renewtheorem{proof}{Proof}

\usepackage{listings}

\newif\ifelec

\electrue

\lstdefinelanguage{SSR}{
mathescape=true,
texcl=false,
morekeywords=[1]{
Section, Module, End, Require, Import, Export, Defensive, Function, Axioms,
Variable, Variables, Parameter, Parameters, Axiom, Hypothesis, Hypotheses,
Notation, Local, Tactic, Reserved, Scope, Open, Close, Bind, Delimit,
Program, Definition, Let, Ltac, Fixpoint, CoFixpoint, Add, Morphism, Relation,
Implicit, Arguments, Unset, Contextual, Strict, Prenex, Implicits,
Inductive, CoInductive, Record, Structure, Canonical, Coercion,
Theorem, Lemma, Corollary, Proposition, Fact, Remark, Example,
Proof, Goal, Save, Qed, Defined, Hint, Resolve, Rewrite, View,
Search, SearchAbout, Show, Print, Printing, All, Graph, Projections, inside,
outside, Locate, Maximal, Eval, Compute, Time, Check, Print, About,
Inline, Class, Instance, Context, Include, Declare, Extraction},
morekeywords=[2]{forall, fun, fix, cofix, struct,
      match, with, end, as, in, return, let, if, is, then, else,
      for, of, nosimpl, where, True, False, beta, delta, zeta, iota},
morekeywords=[3]{Set, Type, Prop},
morekeywords=[4]{
         exists, exists2,
         pose, set, move, case, elim, apply, clear,
            hnf, intro, intros, generalize, rename, pattern, after,
	    destruct, induction, using, refine, inversion, injection,
            constructor,
         rewrite, congr, unlock, compute, vm_compute, native_compute,
            replace, fold, unfold, change, cutrewrite, simpl,
            cbv, lazy,
         have, suff, wlog, suffices, without, loss, nat_norm,
            assert, cut, trivial, revert, bool_congr, nat_congr,
	 symmetry, transitivity, auto, split, left, right,
         autorewrite,
       interval_intro},
morekeywords=[5]{
         by, done, exact, reflexivity, tauto, firstorder, romega, omega,
         assumption, solve, contradiction, discriminate,
         ring, field, interval},
morekeywords=[6]{do, last, first, try, idtac, repeat, progress},
literate=
        {:=}{{$\mathrel{\mathop:\mathopen=}$}}2
        {=}{{$=$}}1
        {==}{{$\equiv$}}1
        {============================}{{============================}}9
        {!=}{{$\not\equiv$}}1
        {<=}{{$\leq$}}1
        {>=}{{$\geq$}}1
        {<>}{{$\neq$}}1
        {->}{{$\rightarrow$}}2
        {<-}{{$\leftarrow$}}2
        {=>}{{$\Rightarrow$}}1
        {/\\}{{$\wedge$}}2
        {\\/}{{$\vee$}}2
        {<->}{{$\leftrightarrow$}}2
        {<=>}{{$\Leftrightarrow$}}2
        {forall\ }{{$\forall\,$}}1
        {exists\ }{{$\exists\,$}}1
        {negb}{{$\neg$}}1
        {~}{{$\neg$}}1
        {\\in}{{$\in$}}1
        {^-1}{{$^{-1}$}}1
  {á}{{\'a}}1 {é}{{\'e}}1 {í}{{\'i}}1 {ó}{{\'o}}1 {ú}{{\'u}}1
  {Á}{{\'A}}1 {É}{{\'E}}1 {Í}{{\'I}}1 {Ó}{{\'O}}1 {Ú}{{\'U}}1
  {à}{{\`a}}1 {è}{{\`e}}1 {ì}{{\`i}}1 {ò}{{\`o}}1 {ù}{{\`u}}1
  {À}{{\`A}}1 {È}{{\'E}}1 {Ì}{{\`I}}1 {Ò}{{\`O}}1 {Ù}{{\`U}}1
  {ä}{{\"a}}1 {ë}{{\"e}}1 {ï}{{\"i}}1 {ö}{{\"o}}1 {ü}{{\"u}}1
  {Ä}{{\"A}}1 {Ë}{{\"E}}1 {Ï}{{\"I}}1 {Ö}{{\"O}}1 {Ü}{{\"U}}1
  {â}{{\^a}}1 {ê}{{\^e}}1 {î}{{\^i}}1 {ô}{{\^o}}1 {û}{{\^u}}1
  {Â}{{\^A}}1 {Ê}{{\^E}}1 {Î}{{\^I}}1 {Ô}{{\^O}}1 {Û}{{\^U}}1
  {œ}{{\oe}}1 {Œ}{{\OE}}1 {æ}{{\ae}}1 {Æ}{{\AE}}1 {ß}{{\ss}}1
  {ű}{{\H{u}}}1 {Ű}{{\H{U}}}1 {ő}{{\H{o}}}1 {Ő}{{\H{O}}}1
  {ç}{{\c c}}1 {Ç}{{\c C}}1 {ø}{{\o}}1 {å}{{\r a}}1 {Å}{{\r A}}1
  {€}{{\euro}}1 {£}{{\pounds}}1 {«}{{\guillemotleft}}1
  {»}{{\guillemotright}}1 {ñ}{{\~n}}1 {Ñ}{{\~N}}1 {¿}{{?`}}1,
comment=[s]{(*}{*)},
showstringspaces=false,
morestring=[b]",
morestring=[d]´,
extendedchars=true,
sensitive=true,
breaklines=true,
basicstyle=\ttfamily,
columns=[l]fullflexible,
keepspaces=true,
identifierstyle={\ttfamily\ifelec\color{black}\fi},
keywordstyle=[1]{\bfseries\ttfamily\ifelec\color{dkviolet}\fi},
keywordstyle=[2]{\ttfamily\ifelec\color{dkgreen}\fi},
keywordstyle=[3]{\ttfamily\ifelec\color{dkgreen}\fi},
keywordstyle=[4]{\ttfamily\ifelec\color{dkblue}\fi},
keywordstyle=[5]{\ttfamily\ifelec\color{red}\fi},
keywordstyle=[6]{\ttfamily\ifelec\color{dkpink}\fi},
stringstyle=\ttfamily,
commentstyle={\ttfamily\ifelec\color{firebrick}\fi},
moredelim=[is][\bfseries\ttfamily\ifelec\color{red}\fi\underbar]{|*}{*|},
moredelim=*[is][\itshape\rmfamily]{/*}{*/},
moredelim=[is][\bfseries\ttfamily\ifelec\color{dkviolet}\fi]{\{-}{-\}}
}

\definecolor{dkblue}{rgb}{0,0.1,0.5}
\definecolor{lightblue}{rgb}{0,0.5,0.5}
\definecolor{dkgreen}{rgb}{0,0.4,0}
\definecolor{dk2green}{rgb}{0.4,0,0}
\definecolor{dkviolet}{rgb}{0.6,0,0.8}
\definecolor{dkpink}{rgb}{0.75,0,1}
\definecolor{firebrick}{rgb}{0.69,0.13,0.13}

  {\begin{Sbox}\begin{minipage}{0.97\textwidth}%
    \begin{flushright}\textcolor{red}{\fbox{Version 1.3}}%
      \end{flushright}\noindent}%
  {\end{minipage}\end{Sbox}\noindent\doublebox{\TheSbox}}

\newcommand{\coq}{\lstinline[language=SSR,mathescape=true]}

\newcommand{\lstsetup}[2]{%
\lstset{#1,%
  columns=[l]fullflexible,%
  basicstyle=\ttfamily,%
  mathescape=true,
  escapechar=§,%
  tabsize=2,
  breaklines=true,
  xleftmargin=2.3mm,%
  #2}}

\definecolor{grey-code}{rgb}{0.85,0.85,0.86}

\lstnewenvironment{coqcode}[1][]%
{\noindent\ignorespaces\lstsetup{backgroundcolor=\color{grey-code},language=SSR}{xleftmargin=0mm,basicstyle=\ttfamily\small,#1}}%
{}

\newcommand{\kw}[1]{\textit{#1}}
\newcommand{\precP }{\prec^{\mathcal{P}}}
\newcommand{\NE}{\bf}
\newcommand{\prefone}{\mathit{pref1}}
\newcommand{\preftwo}{\mathit{pref2}}
\newcommand{\gf}{\mathit{gf}}
\newcommand{\false}{\mathit{false}}
\newcommand{\true}{\mathit{true}}

\title{An Existence Theorem of Nash Equilibrium\\
 in Coq and Isabelle\footnote{This work was partly supported by the FAGames project of LabEx CIMI.~~S.~Le~Roux was also partly supported by the ERC inVEST (279499) project.}}
\author{Stéphane Le Roux
\institute{Université Libre de Bruxelles}
  \email{stephane.le.roux@ulb.ac.be}
\and
Érik Martin-Dorel
\institute{IRIT, Université de Toulouse}
\email{\quad erik.martin-dorel@irit.fr}
\and
Jan-Georg Smaus
\institute{IRIT, Université de Toulouse}
\email{jan-georg.smaus@irit.fr}
}

\def\titlerunning{Nash equilibrium in
  Coq and Isabelle}
\def\authorrunning{S. Le Roux and É. Martin-Dorel and J.-G. Smaus}
\begin{document}

\maketitle

\begin{abstract}
  Nash equilibrium (NE) is a central concept in game theory. Here we
  prove formally a published theorem on existence of an NE in two proof
  assistants, Coq and Isabelle: starting from a game with finitely many outcomes, one may derive a game by
rewriting each of these outcomes with either of two basic outcomes,
namely that Player~$1$ wins or that Player~$2$ wins. If all ways of
deriving such a \emph{win/lose game} lead to a game where one player
has a winning strategy, the original game also has a Nash
equilibrium.

  This article makes three other contributions: first, while the
  original proof invoked linear extension of strict partial orders,
  here we avoid it by generalizing the relevant definition. Second, we
  notice that the theorem also implies the existence of a secure
  equilibrium, a stronger version of NE that was introduced for model
  checking. Third, we also notice that the constructive proof of the
  theorem computes secure equilibria for non-zero-sum priority games
  (generalizing parity games) in quasi-polynomial time.
\end{abstract}

\section{Introduction and motivations}
\label{intro-sec}

The four-color theorem was the first major theorem proved with the
assistance of a computer in 1976. In~\cite{1976} the computer was
merely checking thousands of cases via a dedicated program, thus
completing an otherwise paper-and-pencil proof. In~\cite{Gonthier2008}
the computer checked the whole formalized proof via a general-purpose
and widely used software named Coq. Since then, other challenging (or
just interesting) theorems have been likewise formalized and checked.

In 2006 Coq was used by~\cite{VESTERGAARD200646}, which was
generalized by~\cite{LeRoux2009}, to formalize and check a result from
game theory: Kuhn's existence of a Nash equilibrium (NE) in finite
games in extensive form. Coq was also used to deal with some infinite
games in extensive form~\cite{Lescanne2012}, or with random Boolean
games~\cite{MDS2017}. Isabelle/HOL, another proof software, was used
recently~\cite{Dittmann2016} to formalize and check a result in game
theory for logic and computer science: the positional determinacy of
parity games.

This article formalizes a game-theoretic result, essentially~\cite[Lemma
2.4]{SLR2014},
both in Coq and Isabelle. The result is as follows: starting
from a game with finitely many outcomes, one may derive a game by
rewriting each of these outcomes with either of two basic outcomes,
namely that Player~$1$ wins or that Player~$2$ wins. If all ways of
deriving such a \emph{win/lose game} lead to a game where one player
has a winning strategy, the original game also has an NE. We chose 
to prove this result for several reasons:

\begin{itemize}
\item It lies at the boundary of traditional game theory and game
  theory for logic and computer science. Indeed, it extends
  determinacy, typically a logic and computer science concern, into existence of
  NE, typically a game-theoretic concern.
  As examples, \cite[Lemma 2.4]{SLR2014} 
  generalizes Borel determinacy~\cite{Martin75}, finite-memory
  determinacy of Muller games~\cite{GH82}, and positional determinacy
  of parity games~\cite{EJ91}. In this article, we further notice that
  the theorem also implies the existence of a secure equilibrium, a 
  stronger version of NE that was introduced in~\cite{CHATTERJEE200667} for model checking.

\item The proof is constructive and the corresponding algorithm
  (building an NE provided that we can solve determinacy) has linear
  time complexity in the number of outcomes. Since a recent
  breakthrough~\cite{CJKLS2017} shows that parity games can be solved
  in quasi-polynomial time, we note in this article that secure
  equilibria for non-zero-sum priority games (generalizing parity
  games) can also be computed in quasi-polytime.

\item The result features games in normal form, a very general class of games that includes finite and infinite games in extensive form discussed in~\cite{VESTERGAARD200646},~\cite{LeRoux2009},~\cite{Lescanne2012} and~\cite{Martin75}, and Muller and parity games discussed in~\cite{GH82},~\cite{EJ91},~\cite{Dittmann2016}, and~\cite{CJKLS2017}.

\item The result's statement and proof involve and combine several basic concepts in game theory, so the ability to handle them properly could constitute a basis for a usable game-theory library.

\item The result is a slight weakening of \cite[Lemma 2.4]{SLR2014}, i.e. the interesting part of the lemma. The full lemma is only technically needed in \cite{SLR2014}, since it is the base case of the big proof of \cite[Theorem 2.7]{SLR2014}. The big proof goes by induction on the order types of the inverses of the players' preferences, where the order types are assumed to be countable ordinals. So, this article essentially proves the base case of the induction and leaves the inductive case for future work.

\item Variants of the base case were proved independently
  in~\cite{Gurvich89} and~\cite{Gurvich75}. 
The fact that the idea behind the theorem emerged in different communities suggests that it is broadly interesting.

\end{itemize}

A significant contribution of this article is the modification of the
proof structure of~\cite[Lemma 2.4]{SLR2014} to simplify its
formalization: In~\cite[Lemma 2.4]{SLR2014}, the preferences
are 
extended linearly in the beginning of the proof. Then the new
linear preferences are lifted to subsets of outcomes, where the
definition of the lift hinges upon the linearity assumption. This
helps find an NE for the new preferences, which is also an NE for the
original ones. While it is convenient to invoke linear extension in
the paper-and-pencil proof, it is costly to formalize. 
It was
already formalized in Coq in~\cite{LeRoux2009} (and improved
in~\cite{alglave:inria-00604656} in terms of algorithmic complexity),
but we prefer to avoid relying too much on external libraries. So
we generalize the lift such that the
input may be an arbitrary partial order instead of 
necessarily a linear order.

\paragraph{Organization of the paper:} Section~\ref{sect:def} gives background definitions in game
theory; Section~\ref{sec:lift-order} 
generalizes the lift of the
preference; Section~\ref{sec:pen-paper} gives a new paper-and-pencil
proof of~\cite[Lemma 2.4]{SLR2014} without invoking linear extension; Section~\ref{sec:ec-se} discusses secure equilibria and their computation; Sections~\ref{sec:coq} and~\ref{sec:isabelle} describe the formal proofs in Coq and
Isabelle/HOL, respectively. Section~\ref{sec:concl} gives concluding remarks.

\noindent
The formal developments are available at
\url{https://www.irit.fr/~Erik.Martin-Dorel/equi-thm/}.

\section{Background definitions}\label{sect:def}

Game forms (introduced in~\cite{Gibbard73}) are the central concept of our article. They can be instantiated into games by providing preferences for the players. Then, the Nash equilibria are defined for games. The win/lose games and their winning strategies are an important special case. We recall all this below.

\begin{definition}\label{defn:gf}
A \kw{game form} is a tuple $\langle A,(S_a)_{a\in A},O,v\rangle$ such that 
\begin{itemize}
\item $A$ is a nonempty set (of players, or agents),
\item $\prod_{a\in A}S_a$ 
is a nonempty Cartesian product (whose elements are the strategy profiles and where $S_a$ represents the strategies available to player $a$),
\item $O$ is a nonempty set (of possible outcomes),
\item $v:\prod_{a\in A} S_a\to O$ is the outcome function that values the strategy profiles.
\end{itemize}
A game form endowed with a binary relation $\prec_a$ over $O$ for each player $a$
(modeling her preference) is called a \kw{game in normal form}. 
In the remainder we just write ``game'' for short.
\end{definition}
\begin{definition}[Nash equilibrium]\label{defn:ne}
Let $\langle A,(S_a)_{a\in A} ,O,v,(\prec_a)_{a\in A}\rangle$ be a game. A strategy
profile $s$ in $S:=\prod_{a\in A} S_a$ is a \kw{Nash equilibrium} if it
makes every player $a$ stable, i.e., $v(s)\not\prec_a v(s')$ for all $s'\in
S$ that differ from $s$ at most in the $a$-component:
\[NE(s)\quad:=\quad\forall a\in A,\forall s'\in S,\quad\left(\forall b\in A\setminus\{a\},\,s_b= s'_b\right)~\Rightarrow~v(s)\not\prec_a v(s')\]
\end{definition}
Four games are %
shown in Figure~\ref{fig:four-games}, with Players
$1$ and $2$ who have two strategies each. Here, the outcomes are in
$\mathbb{R}^2$ and are called \kw{payoff pairs}. Player~$1$ ($2$)
prefers payoff pairs with greater first (second) component. In the
first game, if Player~$1$ picks the strategy $1_t$ and Player~$2$
picks $2_l$, the strategy profile $(1_t,2_l)$ then yields payoff $1$
for Player~$1$ and $0$ for Player~$2$.  The payoff pairs that
correspond to NEs are written in bold. E.g., the second game has
no NE. Note that the usual definition of NE uses $\geq$ to compare real 
numbers, but in our general setting, using $\not\prec_a$ instead expresses
exactly the intended concept of NE.
\begin{figure}[h]
  \centering
\begin{displaymath}
\begin{array}{c@{\hspace{0.8cm}}c@{\hspace{0.8cm}}c@{\hspace{0.8cm}}c}
\begin{array}{c|c@{,\;}c@{\;\vline\;}c@{,\;}c|}
          \multicolumn{1}{c}{}&
	  \multicolumn{2}{c}{2_l}&
	  \multicolumn{2}{c}{2_r}\\
	  \cline{2-5}
 	  1_t & 1 & 0 & \NE 5 & \NE 0 \\
	  \cline{2-5}
	  1_b & \NE 2 & \NE 4 & 5 & 3\\
	  \cline{2-5}
\end{array}
&
\begin{array}{c|c@{,\;}c@{\;\vline\;}c@{,\;}c|}
          \multicolumn{1}{c}{}&
	  \multicolumn{2}{c}{2_l}&
	  \multicolumn{2}{c}{2_r}\\
	  \cline{2-5}
 	  1_t & 0 & 1 & 1 & 0 \\
	  \cline{2-5}
	  1_b & 1 & 0 & 0 & 1\\
	  \cline{2-5}
	 \end{array}
&
\begin{array}{c|c@{,\;}c@{\;\vline\;}c@{,\;}c|}
          \multicolumn{1}{c}{}&
	  \multicolumn{2}{c}{2_l}&
	  \multicolumn{2}{c}{2_r}\\
	  \cline{2-5}
 	  1_t & \NE 2 & \NE 1 & 0 & 0 \\
	  \cline{2-5}
	  1_b & 0 & 0 & \NE 1 & \NE 2\\
	  \cline{2-5}
	 \end{array}
&
\begin{array}{c|c@{,\;}c@{\;\vline\;}c@{,\;}c|}
          \multicolumn{1}{c}{}&
	  \multicolumn{2}{c}{2_l}&
	  \multicolumn{2}{c}{2_r}\\
	  \cline{2-5}
 	  1_t & 0 & 1 & 0 & 1 \\
	  \cline{2-5}
	  1_b & \NE 1 & \NE 0 & \NE 1 & \NE 0\\
	  \cline{2-5}
\end{array}
\end{array}
\end{displaymath}
\caption{Four two-player games with two strategies each}
\label{fig:four-games}
\end{figure}

\begin{definition}\label{def:win-lose}
\begin{itemize}
\item A \kw{win/lose game} is a game where $A=\{1,2\}$ and
  $O=\{(1,0),(0,1)\}$ and the preferences are defined by $(0,1)\prec_1(1,0)$
  and $(1,0)\prec_2(0,1)$. 

\item A \kw{winning strategy} for Player~$1$ is a strategy $s_1\in S_1$ such that $v(s_1,s_2)=(1,0)$ for all $s_2\in S_2$. A winning strategy for Player~$2$ is a strategy $s_2\in S_2$ such that $v(s_1,s_2)=(0,1)$ for all $s_1\in S_1$.
\item A win/lose game such that one player has a winning strategy is
  said to be \kw{determined}. 
For $i\in\{1,2\}$, if Player~$i$ is the winning player and 
the winning strategy is in some $R_i \subseteq S_i$, the game is said to be 
\kw{determined via} $R_i$.

\end{itemize}
\end{definition}

The second game of Figure~\ref{fig:four-games}  is a non-determined win/lose game. The fourth game is also win/lose, and $1_b$ is a winning strategy for Player~$1$, so the game is determined.

The notion of winning strategy is relevant for win/lose games only, but the following remark clarifies why the transfer from winning strategy to multi-outcome Nash equilibrium is a process of generalization.

\begin{remark}\label{rmk:ws-ne} 
A win/lose game has a winning strategy iff it has a Nash equilibrium.
\end{remark}

Remark~\ref{rmk:ws-ne} is formalized as the Coq lemma
\coq{determined_iff_NE} in Section~\ref{sec:coq-ws-ne},
and its right-to-left implication is also formalized as the Isabelle theorem \texttt{\textsl{someone\_wins}} in Section~\ref{sec:ws-ne}.

From a two-player game form one may derive both games and win/lose games:

\begin{definition}\label{defn:is-dg-ds} Let $G = \langle \{1,2\},S_1,S_2,O,v\rangle$ be a two-player game form.
\begin{enumerate}
\item For all $\prec_1,\prec_2 \subseteq O^2$ the game $\langle \{1,2\},S_1,S_2,O,v,\{\prec_1,\prec_2\}\rangle$ is said to be \kw{derived} from $G$.
\item Let $wl: O \to \{(1,0),(0,1)\}$. The win/lose game $\langle S_1,S_2,wl\circ v\rangle$ is also said to be \kw{derived} from $G$. 
\item 
\label{determined-gameform-item}
Let $R_1\subseteq S_1$ and $R_2\subseteq S_2$. If all win/lose games derived
  from a game form are determined (via  
$R_1$ or $R_2$, resp., depending on who wins), 
the game form is also said to be \kw{determined} (\kw{via} %
$R_1$ \kw{and} $R_2$).

\item\label{defn:is-dg-ds4} Let $P\subseteq O$, and let $s_1\in S_1$ be such that $v(s_1,S_2):=\{v(s_1,s_2)\,\mid\,s_2\in S_2\}\subseteq P$. The strategy $s_1$ is said to \kw{enforce} $P$. (And likewise $s_2 \in S_2$ may enforce subsets of outcomes.)
\end{enumerate}
\end{definition}

The concept of \emph{determined game form} is used to state our theorem. 
The leftmost game form in Figure~\ref{fig:four-game-forms} is not
determined, e.g.,
because %
instantiating $X$ with $(0,1)$ and $Y$ with $(1,0)$ yields a
non-determined game, namely the second game in
Figure~\ref{fig:four-games}. The second game form in
Figure~\ref{fig:four-game-forms} is not determined either. The third game form in
Figure~\ref{fig:four-game-forms} is determined: if $Y$ is mapped to
$(1,0)$ then $1_b$ is winning for Player~$1$; if $X$ and $Z$ are
mapped to $(1,0)$ then $1_t$ is winning for Player~$1$; else either
$2_l$ or $2_r$ is winning for Player~$2$. The last game form in
Figure~\ref{fig:four-game-forms} is also determined: if $Y$ is mapped
to $(1,0)$ then $1_b$ is winning for Player~$1$; else $2_r$ is winning
for Player~$2$. Note that the winner of a game obtained by
instantiating a determined game form may depend on the instance.
\begin{figure}[h]
  \centering
\begin{displaymath}
\begin{array}{c@{\hspace{1cm}}c@{\hspace{1cm}}c@{\hspace{1cm}}c}
\begin{array}{c|c@{\;\vline\;}c|}
          \multicolumn{1}{c}{}&
	  \multicolumn{1}{c}{2_{l}}&
	  \multicolumn{1}{c}{2_{r}}\\
	  \cline{2-3}
 	  1_{t} & X & Y \\
	  \cline{2-3}
	  1_{b} & Y & X\\
	  \cline{2-3}
\end{array}
&
\begin{array}{c|c@{\;\vline\;}c@{\;\vline\;}c|}
	  \multicolumn{1}{c}{}&
	  \multicolumn{1}{c}{2_{l}}&
	  \multicolumn{1}{c}{2_{m}}&
	  \multicolumn{1}{c}{2_{r}}\\
       	  \cline{2-4}
 	  1_t & X & Y &Z \\
	  \cline{2-4}
	  1_b & Y & Z &X\\
	  \cline{2-4}
\end{array}
&
\begin{array}{c|c@{\;\vline\;}c|}
	  \multicolumn{1}{c}{}&
	  \multicolumn{1}{c}{2_{l}}&
	  \multicolumn{1}{c}{2_{r}}\\
       	  \cline{2-3}
 	  1_t & X & Z\\
	  \cline{2-3}
	  1_b & Y & Y\\
	  \cline{2-3}
\end{array}
&
\begin{array}{c|c@{\;\vline\;}c@{\;\vline\;}c|}
	  \multicolumn{1}{c}{}&
	  \multicolumn{1}{c}{2_{l}}&
	  \multicolumn{1}{c}{2_{m}}&
	  \multicolumn{1}{c}{2_{r}}\\
       	  \cline{2-4}
 	  1_t & X & Z &Y \\
	  \cline{2-4}
	  1_b & Y & Y &Y\\
	  \cline{2-4}
\end{array}
\end{array}
\end{displaymath}
  \caption{Four two-player game forms}
  \label{fig:four-game-forms}
\end{figure}

Note that since $S_1$ and $S_2$ are nonempty by definition, no player can enforce the empty set.

Also note that the subsets $R_i$ from Definition~\ref{defn:is-dg-ds} represent
strategies of special interest. For instance, Muller games are
determined \textit{via} strategies that can be described by finite
automata~\cite{GH82}, and parity games are determined \textit{via} strategies that are
called \emph{positional}~\cite{EJ91}.

Finally note that, given a two-player game form $G$, Player~$a$ can enforce an outcome subset $P$  in $G$ iff $P \in E_G(\{a\})$ where $E_G$ is the
\emph{effectivity function} of $G$.  These functions were introduced in
\cite{MOULIN1982115} and are now widely used in cooperative game
theory and social choice. Here we use only a special case of them.

Lemma~\ref{lem:d-e} connects
Definitions~\ref{defn:is-dg-ds}.\ref{determined-gameform-item}
and~\ref{defn:is-dg-ds}.\ref{defn:is-dg-ds4}. It is key in the
original and in the formalized proofs.

\begin{lemma}\label{lem:d-e} 
A game form is determined (via $R_1$ and $R_2$) iff for each subset of the outcomes,
either the subset can be enforced by Player~$1$ (via $R_1$), or its complement
can be enforced by Player~$2$ (via $R_2$).
\end{lemma}

Note that in the lemma one can interchange $1$ and $2$ and obtain an
equivalent statement, but one cannot replace ``by Player \ldots'' with the
phrase ``by one of the players''! See the second game form in 
Figure~\ref{fig:four-game-forms}: for every subset $P\subseteq\{X, Y, Z\}$,
either $P$ or $\{X, Y, Z\}\setminus P$ can be enforced by one of the players, but
$\{X,Y\}$, e.g., cannot be enforced by Player~$1$ and its complement
cannot be enforced by Player~$2$.

\section{Lifting the preference without prior linear extension}
\label{sec:lift-order}

We now present the extension of the lift of the preference, which we
mentioned in the end of Section~\ref{intro-sec}. In~\cite{SLR2014} the
lift required that the preference be a linear order; below we extend
the definition to arbitrary partial orders. In the whole section, $\prec$
is a strict partial order over a set $O$, i.e., it is a
binary relation that is transitive and irreflexive. (Irreflexivity means that it satisfies $\forall x \in O, ¬(x \prec x)$.)

\begin{definition}\label{defn:set-pref}
\begin{itemize}
\item For all $x \in O$, let $u(x) := \{z\in O\,\mid\, x \prec z\}$, the \kw{strict upper set} of $x$.
\item For all $Y \subseteq O$, let $u(Y) := \cup_{y\in Y }u(y)$, the \kw{strict upper set} of $Y$.
\item For $A,B \subseteq O$, let $A\precP\!B\,$ (\kw{lift} of $\prec$) iff $\exists A' \subseteq A\setminus B,~~ A'\neq\emptyset ~\land~ A \setminus (A' \cup u(A')) = B \setminus (A' \cup u(A'))\bigr)$.
\end{itemize}
\end{definition}

Let us give some intuition behind Definition~\ref{defn:set-pref}.
First, $\precP $ is irreflexive since there is no nonempty $A'$ in
$A\setminus A$. Second, if $B \subsetneq A \subseteq O$, 
picking $A' := A \setminus B$ shows
that $A \precP B$. In particular, $\emptyset$ is the only $\precP$-maximal set
and $O$ the only $\precP$-minimal
set. %
More generally, note that for all sets $A,B,C$ we have $A \setminus C = B \setminus C$
iff $A\Delta B \subseteq C$, where $A\Delta B := (A\setminus B) \cup (B \setminus A)$ is the symmetric
difference of $A$ and $B$. Therefore $A\precP B$ iff there is a
nonempty $A' \subseteq A \setminus B$ such that $A \Delta B \subseteq A' \cup u(A')$. (This
emphasizes that whether $A\precP B$ only depends on how $\prec$ behaves on
$A\Delta B$, the points where $A$ and $B$ disagree.)

Let us further assume 
that $O$ is finite in the remainder of the article, so $A\precP B$ iff the
minimal elements of $A \Delta B$ are all in $A \setminus B$. It is now easy to see
that for all $A,B \subseteq O$, if $A \precP B$, the $A'$ witnessing it are
exactly the subsets of $A \setminus B$ containing all the minima of $A \setminus B$
(and thus of $A \Delta B$).

Note that although the lift of a preference coincides with the preference on singleton outcomes, it is difficult to interpret the full lift game-theoretically. %
This remark holds even for linear
preferences, as shown by considering the usual order $<$ on the natural numbers. Then $\{1,2,3,4,5\}\ <^{\mathcal{P}} \{1\} <^{\mathcal{P}} \{2,3,4,5\} <^{\mathcal{P}} \{2,3\} <^{\mathcal{P}} \{2,4,5\} <^{\mathcal{P}} \{2\} <^{\mathcal{P}} \{3,5\}<^{\mathcal{P}} \{3\} <^{\mathcal{P}} \{4\} <^{\mathcal{P}} \emptyset$.

If $O$ is finite, the characterization of $\precP $ using the minima of the symmetric difference is more intuitive than Definition~\ref{defn:set-pref} itself, yet it is unclear whether it is easier to handle with a proof assistant.

Lemma~\ref{lem:set-pref-po} proves that $\precP $ is a strict partial order, just like $\prec$. Note that dropping the non-emptiness condition (of $A'$) in Definition~\ref{defn:set-pref} would yield a partial order, i.e., the reflexive closure of the actual $\precP $.

\begin{lemma}%
\label{lem:set-pref-po}
If $O$ is finite, $\precP $ is a strict partial order.

\begin{proof}
It is irreflexive as discussed above; the main difficulty is to prove transitivity. Let $A \precP B$ be witnessed by $A'\neq \emptyset$ and $B \precP C$ be witnessed by $B' \neq \emptyset$. Since $A'\subseteq O$, $B'\subseteq O$, and $O$ is finite, $A'\cup B'$ is finite too. Let $A''$ be the minimal elements of $A' \cup B'$. We argue below that $A \precP C$ is witnessed by $A''$.

Let $x\in A''$, and let us first prove that $x\notin C$. First case, $x\in B'$, so $x\notin C$ by definition. Second case, $x\in A'$, so $x\in A$ but $x\notin B$ and in particular $x\notin B'$. For all $y \prec x$ we have $y \notin B'$ by definition of $A''$, so $x\notin u(B')$, so $x\notin B' \cup u(B')$. By definition $B \setminus (B' \cup u(B')) = C \setminus (B' \cup u(B'))$, so $x\notin C$ since $x\notin B$. Let us now prove that $x \in A$. First case, $x \in A'$, so $x \in A$ by definition. Second case, $x\in B'$, so $x\in B$, and $x\notin A'$. Moreover $y \notin A'$ for all $y \prec x$ by definition of $A''$, so $x \notin A' \cup u(A')$. By definition $A \setminus (A' \cup u(A')) = B \setminus (A' \cup u(A'))$, so $x\in A$ since $x\in B$.

Let us finally prove that $A \setminus (A'' \cup u(A'')) = C \setminus (A'' \cup u(A''))$. Let $x\notin A'' \cup u(A'')$, so $x\notin A' \cup u(A')$ and $x\notin B'\cup u(B')$ since $ A' \cup u(A') \cup B'\cup u(B') \subseteq A'' \cup u(A'')$. (Equality holds but is not needed here.) Therefore $x \in A$ iff $x\in B$ (since $x\notin A' \cup u(A')$) and $x \in B$ iff $x\in C$ (since $x\notin B' \cup u(B')$).
\end{proof}
\end{lemma}

\noindent Note that if $\prec$ and set membership are decidable in polynomial time in the cardinality of $O$, so is deciding $A \precP  B$: decide whether $A \neq B$ and whether for all elements in $B \setminus A$ there is a smaller element in $A \setminus B$.

Beside being convenient for our proofs, Definition~\ref{defn:set-pref} might be useful outside of game theory. It provides a canonical way to lift a finite order of elements to a finite order of subsets. We can already note that it can be generalized to finite multisets: for all multisets $f,g: O \to \mathbb{N}$, let us define $f \precP  g$ iff $\{x \in O \mid g(x) < f(x)\} \precP  \{x \in O \mid f(x) < g(x)\}$.

\section{Paper-and-pencil proof of \cite[Lemma 2.4]{SLR2014} using 
Lemma~\ref{lem:set-pref-po}} %
\label{sec:pen-paper}

Let us now state the theorem and provide the alternative paper-and-pencil proof that is easier to formalize. We recall that the original proof invokes linear extension of the preferences, whereas our new proof uses only 
Lemma~\ref{lem:set-pref-po}.

\begin{theorem}[Finitary equilibrium transfer]\label{thm:fet}
Let $\langle \{1,2\}, S_1,S_2,O,v,\{\prec_1,\prec_2\}\rangle$ be a two-player game with finite $O$, 
let $R_1\subseteq S_1$ and $R_2\subseteq S_2$ be subsets of the strategy sets,
and let us assume the following:
\begin{enumerate}
\item\label{cond:lem-et1} the underlying game form is determined
  via $R_1$ and $R_2$;
\item\label{cond:lem-et2} both preferences $\prec_1$ and $\prec_2$ are strict partial orders.%

\end{enumerate}
Then the game $\langle \{1,2\}, S_1,S_2,O,v,\{\prec_1,\prec_2\}\rangle$ has a Nash equilibrium in $R_1× R_2$.
\begin{proof}
We number the paragraphs of the proof to facilitate comparison with the formal proofs.

\begin{enumerate}

\item\label{M-item} 
Let $\prec_1^{\mathcal{P}}$ be the lift of $\prec_1$ along
Definition~\ref{defn:set-pref}. It is a strict partial order by
Lemma~\ref{lem:set-pref-po}. Let $M$ be a
$\prec_1^{\mathcal{P}}$-maximal subset of $O$ that Player~$1$ can
enforce via $R_1$ and let $s_1\in R_1$ be a strategy enforcing $M$.
\item\label{m-item}
$M$ is finite, as a subset of $O$, and nonempty, since no player can enforce the empty set. Since $\prec_2$ is a strict
partial order, let $m$ be $\prec_2$-maximal in $M$, 
and let $M' := (M \setminus \{m\}) \cup u(m)$ (where $u(m)$ denotes $\{z\in M'\,\mid\, m \prec_1 z\}$, the strict
upper set with respect to Player~$1$'s preference). 
Note that for all sets $A,B,C$ such that $B \subseteq C$ we have $(A \setminus B)\setminus C =  (A
\cup B)\setminus C = A \setminus C$. So $M \setminus (\{m\} \cup u(m)) = M' \setminus (\{m\} \cup u(m))$. 
Since $m \in M \setminus M'$, it witnesses that $M \prec_1^{\mathcal{P}} M'$. 
\item\label{enforce-item}
By maximality in the definition of $M$, Player~$1$ cannot enforce
$M'$. So Player~$2$ can enforce $O\backslash M'$ by determinacy assumption and
Lemma~\ref{lem:d-e}. Let $s_2\in R_2$ be a strategy enforcing $O\backslash M'$,
so that $v(s_1,s_2)\in M\cap (O\backslash M')=\{m\}$.
\item\label{nash-item}
First, the strategy profile $(s_1,s_2)$ makes Player~$2$ stable,
since $m$ is $\prec_2$-maximal among $M$, which is enforced by $s_1$.
Second, let $o\in O$ be such that $m\prec_1 o$, i.e. $o \in u(m)$,
so $o\in M'$. This shows that $m$ is $\prec_1$-maximal among $O\backslash M'$, which is
enforced by $s_2$. So the profile $(s_1,s_2)$ also makes Player~$1$
stable. Therefore $(s_1,s_2)\in R_1× R_2$ is a Nash equilibrium. \qed
\end{enumerate}
\end{proof}
\end{theorem}
Note that although the roles of the players are symmetric in the
original question of existence of Nash equilibria, the proof of
Theorem~\ref{thm:fet} breaks this symmetry: One player (Player~$1$ in
the proof) first selects a strategy that somehow maximizes her
guarantee in a ``lexicographic-like'' way, and then her opponent
(Player~$2$ in the proof) maximizes his outcome among the available
ones, while locking Player~$1$ into her strategy. (The symmetry need
not be broken if the preferences are antagonistic, though, i.e. if one is the symmetric of the other: both
players may independently pick strategies as Player~$1$ does in the
proof.)
\section{Existence and computation of secure equilibria}\label{sec:ec-se}

The secure equilibria were introduced~\cite{CHATTERJEE200667} in
connection with model checking, for two-player games with outcomes in $\{0,1\}^2$. They were generalized into quantitative secure
equilibria~\cite{Brihaye2010} for two-player games with outcomes in $\mathbb{R}^2$. The (quantitative) secure equilibria of a game
are the NEs 
of another game obtained by changing the usual preference
of each player into malevolent preference: instead of just trying to
maximize her own payoff, she tries primarily to do so and, in case of
ties, to minimize the opponent's payoff. Since these new preferences
are strict linear orders, by Theorem~\ref{thm:fet} we know the following:

\begin{corollary}
If a game form with finitely many outcomes is determined via strategies of some sort, every derived game with real-valued payoffs (and the usual preferences) has a secure equilibrium using strategies of the same sort.
\end{corollary}

Computationally, the hard part of the proof of Theorem~\ref{thm:fet}
is to find $M$ from Paragraph~\ref{M-item}. Potentially there are
indeed exponentially many subsets to check, but it is shown in \cite{SLR2014}
(after Lemma 2.4) that it suffices to check linearly
many of them, by dichotomy. I.e., to compute a Nash or secure
equilibrium it suffices to decide $|O|$ times the winner of a derived
win/lose game, and to compute twice a winning strategy.

Let us now apply the above complexity remark to parity games. A recent breakthrough~\cite{CJKLS2017} shows that deciding the winner and computing a winning strategy can both be done in quasi-polynomial time in parity games. To exploit this, let us (similarly to~\cite[Section 3.2]{SLR2014}) define a priority game by the arena of a parity game and a function from its priorities (in $\mathbb{N}$) to $\mathbb{R}^2$: after an infinite play, the payoff for the first (second) player is the first (second) component of the pair associated with the least priority occurring infinitely often.

\begin{corollary}
Consider priority games with $n$ vertices and $m$ priorities.
\begin{enumerate}
\item Positional secure equilibria can be computed in $O(n^{\log(m) + 7})$.
\item For fixed parameter $m$, there is an FPT-algorithm with runtime $O(n^6) + h(m)$. 
\end{enumerate}
\begin{proof}
In~\cite{CJKLS2017} the numbers are $O(n^{\log(m) + 6})$ and $O(n^5) + g(m)$. By the above complexity remark, they have to be multiplied by $m$, which cannot be greater than $n$ since each vertex carries one priority.
\end{proof} 
\end{corollary}

\section{Coq formal setup for Theorem~\ref{thm:fet}}
\label{sec:coq}
We use the Coq proof assistant along with the SSReflect proof language
and the MathComp
library,\footnote{\url{https://math-comp.github.io/math-comp/}}
especially the following theories: \texttt{fintype} (finite types with
decidable equality), \texttt{finfun} (functions over finite domains),
\texttt{finset} (finite sets), and \texttt{bigop} (iterated
operators). We also use the \texttt{RelationClasses} theory from Coq's
standard library, to facilitate the reasoning on relations that are
not in the scope of MathComp's framework. As of now, the size of the
development is about 1300 lines of Coq code.

\subsection{Main definitions}
\label{sec:coq-def}
We chose to start the formalization by defining
games and Nash equilibrium in the most general way. This general
setting was not compulsory for mechanizing the proofs we focus on, but
it allows one to get a wider game-theory library as a basis for
further developments.

First, we define strategies as follows, relying on the
dependently-typed theory of Coq:
\begin{coqcode}
 Section Generic.
 Variables (Agt : Type) (Strat : Agt -> Type).
 Definition strategy := forall a : Agt, Strat a.
\end{coqcode}
\coq{Agt} and \coq{Strat} represents the space 
(a term more appropriate than ``set'' in Coq)
of agents and
strategies, respectively, and \coq{strategy} represents the
(dependently-typed) space of strategy profiles
(mapping each agent to its space of strategies).

We then define game forms as follows:
\begin{coqcode}
 Variable Outc : Type.
 Record game_form := GameForm
 { preform :> strategy -> Outc ;
   eq_strategy : forall strat' strat, (forall x, strat x = strat' x) ->
     preform strat = preform strat' }.
\end{coqcode}
So a game form is simply defined as a function mapping a strategy
profile to an outcome. Note that the \texttt{eq\_strategy} extensionality property
is required in this Coq setting since \texttt{strategy} is a function
type and equality is not extensional in Coq. (To demonstrate that this
property can be instantiated in practice, we also give another
definition of 2-player game forms as functions of type %
\coq{Strat$_1\,$*$\;$Strat$_2$ -> Outc}, and provide a function
\texttt{game\_form\_of\_alt2} that converts any such function into a
\texttt{game\_form} record.)

We then define games as a game form endowed with a preference
relation --- ``\coq{prefs a o$_1$ o$_2$}'' means that agent \texttt{a} prefers
\texttt{o$_2$} over \texttt{o$_1$}.\footnote{Note that no property is assumed by
  this definition: the preference relation is an arbitrary binary
  relation.}
\begin{coqcode}
 Record game := Game
 { form :> game_form ;
   prefs : Agt -> Outc -> Outc -> bool }.
\end{coqcode}

We then define what it means that a given strategy profile is a Nash
equilibrium (in the Coq code this notion is split into several
definitions, but for the sake of readability we give below a
syntactically equivalent definition, in one go):
\begin{coqcode}
 Definition is_NE (g : game) (strat : strategy) : Prop :=
   forall a : Agt, forall strat' : strategy, (forall b : Agt, a <> b -> strat b = strat' b) ->
   ~ prefs g a (g strat) (g strat').
\end{coqcode}
This means that each agent is stable (he or she has no incentive to
change strategy assuming other agents keep their strategy). We then
introduce a $\Sigma$-type gathering a profile strategy and a proof
that it is an NE:
\begin{coqcode}
 Definition ex_NE (g : game) : Type := {strat : strategy | is_NE g strat}.
\end{coqcode}

Another useful notion is : 
``Player~$a$ can enforce a set $S$ of outcomes (using some strategy
$s_a$)'':
\begin{coqcode}
 Definition can_enforce_by
   (v : game_form) (a : Agt) (S : Outc -> bool) (sa : Strat a) : Prop :=
   forall strat : strategy, strat a = sa -> v strat \in S.
 Definition can_enforce (v : game_form) (a : Agt) (S : Outc -> bool) : Type :=
   {sa : Strat a | can_enforce_by v a S sa}.
 End Generic.
\end{coqcode}

Next, we define the following enumerated types of players and win/lose
outcomes as well as the natural preference relations over these
outcomes:
\begin{coqcode}
 Inductive player := player1 | player2. Inductive winlose_outc := win1 | win2.

 Definition game_form_2 := game_form player. Definition game_2 := game player.

 Definition winlose_prefs (a : player) (o1 o2 : winlose_outc) : bool :=
   match a, o1, o2 with
   | player1, win2, win1
   | player2, win1, win2 => true
   | _, _, _ => false
   end.
\end{coqcode}
We then formalize the notion of winning strategy for a given player
(using dependent types) as well as the notion of determinacy,
following Definition~\ref{def:win-lose}:
\begin{coqcode}
 Definition preferred_outc (a : player) : winlose_outc :=
   if a is player1 then win1 else win2.
 
 Definition win_strat
   (Strat : player -> Type) (v : game_form_2 winlose_outc Strat)
   (a : player) (sa : Strat a) : Prop :=
   forall strat : strategy Strat, strat a = sa -> v strat = preferred_outc a.
 
 Definition determined Strat (v : game_form_2 winlose_outc Strat) : Type :=
   {a : player & {sa : Strat a | win_strat v a sa}}.
\end{coqcode}

Next, we formalize Definition~\ref{defn:is-dg-ds} regarding the
derived win/lose game from a two-player game form, and the notion of
determined form:
\begin{coqcode}
 Program Definition derivedWLGame (Outc : Type) (Strat : player -> Type)
   (wl : Outc -> winlose_outc) (v : game_form_2 Outc Strat)
 : game_2 winlose_outc Strat := Game (GameForm (wl \§§o v) _) winlose_prefs.
 
 Definition determined_form Outc Strat (v : game_form_2 Outc Strat) : Type :=
   forall wl : Outc -> winlose_outc, determined (derivedWLGame wl v).
\end{coqcode}
In previous definitions, it should be noted that the strategy spaces
of all players (declared by variable \coq{Strat : Agt -> Type}) are
arbitrary types.  As a result, formalizing the constraints $R_1
\subseteq S_1$ and $R_2\subseteq S_2$ involved in
Theorem~\ref{thm:fet} cannot be done using MathComp's inclusion of
finite sets. Instead, we introduce another variable %
\coq{(Strat_R : Agt -> Type)} corresponding to
$\left(R_a\right)_{a\in\texttt{Agt}}$ and formalize the inclusion as %
\coq{(forall a : Agt, Strat_R a -> Strat a)}.

We then extend the definitions presented up to now with the condition
``\emph{via} $\prod_{a\in\texttt{Agt}}R_a$''. In particular,
the two predicates below (whose body is omitted for conciseness) are
straightforwardly defined from predicates \coq{is_NE} and
\coq{determined_form}:

\begin{coqcode}
 Definition is_NE_via (Agt Outc : Type) (Strat_R Strat : Agt -> Type) :
   (forall a : Agt, Strat_R a -> Strat a) -> game Outc Strat -> strategy Strat_R -> Prop.
 
 Definition determined_form_via (Outc : Type) (Strat_R Strat : player -> Type) :
   (forall a : player, Strat_R a -> Strat a) -> game_form_2 Outc Strat -> Type.
\end{coqcode}

\subsection{Results and proofs}
\label{sec:coq-proof}

The main result of the Coq formalization is given by the following
theorem, which has been formally verified without relying on any axiom:
\begin{coqcode}
 Theorem finite_equilibrium_transfer :
   forall (Strat : player -> Type) (_ : strategy player Strat)
    $\:$(Outc : finType) (g : game_2 Outc Strat)
    $\:$(Strat_R : player -> Type) (incl : forall a : player, Strat_R a -> Strat a),
   StrictOrder (prefs g player1) ->
   StrictOrder (prefs g player2) ->
   determined_form_via incl (form g) ->
   ex_NE_via incl g.
\end{coqcode}
The first hypothesis of this Coq theorem (\coq{strategy player Strat})
formalizes the requirement that the space of strategy profiles
$\prod_{a\in\texttt{player}} S_a$ is nonempty. This formalized theorem
corresponds precisely to Theorem~\ref{thm:fet}.
This theorem relies on a formal proof of
Lemma~\ref{lem:d-e}, which consists of the following two lemmas:

\begin{coqcode}
 Lemma determined_form_enforce_outc :
   forall (Outc : Type) (Strat : player -> Type) (v : game_form_2 Outc Strat),
   determined_form v <=> (forall S : Outc -> bool,
                           can_enforce v player1 S
                         + can_enforce v player2 (predC S)).
 
 Lemma determined_form_via_enforce_outc :
  forall (Outc : Type) (Strat_R Strat : player -> Type)
   $\:$(incl : forall a : player, Strat_R a -> Strat a)
   $\:$(v : game_form_2 Outc Strat),
   determined_form_via incl v <=> (forall S : Outc -> bool,
                                    can_enforce_via incl v player1 S +
                                    can_enforce_via incl v player2 (predC S)).
\end{coqcode}                                                                  
Here, \coq{(predC S)} denotes the complement of set \coq{S}, and
\coq{+} denotes the disjoint union (analogous to the \coq{\/}
connector, but for \coq{Type} arguments).

\subsection{Confidence lemmas and Remark~\ref{rmk:ws-ne}}
\label{sec:coq-ws-ne}

We have also proven several results that were not needed for
proving the main theorem, but that are helpful to give more intuition
or increase the confidence one can have in the formalized
definitions. For example, regarding \coq{winlose_prefs} and
\coq{preferred_outc}, we have proven:
\begin{coqcode}
 Lemma winlose_prefs_strict : forall a : player, StrictOrder (winlose_prefs a).
 
 Lemma preferred_outc_correct :
   forall (a : player) (o : winlose_outc), ~ winlose_prefs a (preferred_outc a) o.
\end{coqcode}

Then, we have proven the following lemma that is a formal version of
Remark~\ref{rmk:ws-ne}:
\begin{coqcode}
 Lemma determined_iff_NE :
   forall (Strat : player -> Type) (v : game_form_2 winlose_outc Strat),
   determined v * strategy Strat <=> ex_NE (derivedWLGame id v).
\end{coqcode}
Here, symbols \coq{<=>} and \coq{*} are connectors taking \coq{Type}
arguments. They correspond respectively to the usual connectors
\coq{<->} and \coq|/\| (taking \coq{Prop} arguments). The
left-hand-side of this equivalence has two parts: the fact that the
win/lose game is determined, and the fact that the space of strategy
profiles ($\prod_{a\in\texttt{player}}S_a$) is
nonempty.

\section{Isabelle formal setup for Theorem~\ref{thm:fet}}
\label{sec:isabelle}

We use standard Isabelle/HOL in ISAR proof style without any special
libraries. The current proof code has approximately 1100 lines.

\subsection{Main definitions}
Before we define games, Nash equilibrium etc., we need to look at some
technicalities concerning the preference order. Concerning the lifting of the preference order
(Definition~\ref{defn:set-pref}) we have the following Isabelle definitions
for $u$, $u$ on sets, and $\precP$, respectively: 
\Snippet{uconelessP}

\label{recall-O-finite}
Observe that compared to Definition~\ref{defn:set-pref}, there is no
explicit set of outcomes $O$,  
but the type parameter \verb|'a| is used for the type of the outcomes.  
However, it is stated that $A\precP B$ presupposes that $A$ is
finite. We conjecture that one could weaken the requirements even
further by only requiring that the witness set $A'$ is finite -- this
would be a topic for future work. 
We proved in Isabelle that $\precP$ is a strict partial order (Lemma
\ref{lem:set-pref-po}, see also Subsec.~\ref{isabelle-results-subsec}), without the explicit hypothesis 
that $O$ is
finite (which would be difficult to formulate given that we have no
explicit $O$ in Isabelle). Our somewhat weaker implicit hypothesis
that $A$ is finite (see paragraph above) suffices.

In the proof of Theorem~\ref{thm:fet} we do not construct all kinds of pairs $A \precP B$. Instead we only construct a pair by removing a single element from a set $A$ and by replacing it with all the preferred ones w.r.t. a given order $\prec$. This construction is formalized
in Isabelle as follows:

\Snippet{replaceWithPreferred}

Now let us consider game forms and games (Definition~\ref{defn:gf}). In the
current Isabelle formalization, we restrict ourselves to two players, which is sufficient to formalize Theorem~\ref{thm:fet}.
This is in contrast to the Coq formalization of Section~\ref{sec:coq}.
Also, the dependent types of Coq make generic definitions (for an
arbitrary number of players) easier. In the Isabelle formalization, there is nothing to define about
\emph{strategies} and the \emph{outcomes}: they are simply type
parameters.  A \emph{game form} is then given by an outcome function that
maps a pair of strategies to an outcome:

\Snippet{gameform}

\noindent
and a \emph{game} is obtained by adding one preference relation for each of the
two players: 

\Snippet{game}

Then there are functions $\prefone$, $\preftwo$, and $\gf$ that
extract each of the three components of a game in the obvious
way. These are used in the following definition of a Nash equilibrium (Definition~\ref{defn:ne}), i.e., a function
taking a game and two strategies (one per player) and telling whether they
constitute an NE:
\Snippet{isNash}

We now give the definition of a determined (via \ldots) game
(Definition~\ref{def:win-lose}): 

\Snippet{determined}

We now give the definition of the 
\emph{derived win/lose game} (Definition~\ref{defn:is-dg-ds}): 

\Snippet{derivedWLGame}

The function takes as input a game with outcomes of type \verb|'O| and a set
$O$\footnote{In Isabelle the letter ``o''
  has a reserved meaning which is why we used ``ou'' instead.} 
of outcomes, those for which the first player wins. The derived
win/lose game is the game with outcomes of Boolean type\footnote{%
In the Definition~\ref{defn:is-dg-ds}, we assumed that the outcome of a
win/lose game is $(1,0)$ (Player~$1$ gets $1$, Player~$2$ gets $0$) or
$(0,1)$, which may be intuitive, but it is simpler to say the the
outcome is $\true$ (first player wins) or $\false$.}, where the
preference relations say that Player~$1$ says $\false\prec \true$ (expressed
by the $\lambda$-term $\lambda o\;p.p\land¬ o$ 
and vice versa for Player~$2$), and the
outcome function is the original outcome function composed with a
function that says ``$\true$'' for all values in $O$. 

We now give the definition of a determined (via \ldots) game form
(Definition \ref{defn:is-dg-ds}):
\Snippet{determinedForm}

In the paper-and-pencil version, ``Player~$1$ can enforce $P$'' 
(Definition~\ref{defn:is-dg-ds}) is
defined as an overapproximation, i.e., Player~$1$ might even be able to
enforce a subset of $P$. We also have the Isabelle versions of this
notion, but it turned out to be more useful in the formal development
to define an \emph{exact} notion, i.e., exactly the outcomes that may
occur using a given strategy. 
We give the definition for Player~$1$, but there is an analogous definition for Player~$2$:
\Snippet{enforceSet}

\subsection{Results and proofs}
\label{isabelle-results-subsec}
In the presentation of the proof path we choose a top-down approach,
i.e., we present the main result (Theorem~\ref{thm:fet}) and give some
of the lemmas needed to show it.

\Snippet{equilibriumtransferfinite}

The Isabelle version, like the Coq version, corresponds precisely to Theorem~\ref{thm:fet}.

Unlike the Coq version, the Isabelle version does not have an explicit
assumption that the sets of strategies are nonempty. As a matter of
fact, they are formalized as types in Isabelle/HOL (namely, the type
parameters \textsl{\texttt{'S1}, \texttt{'S2}} in the definition of
a \texttt{\textsl{game}}), which are necessarily nonempty.

The proof has 153 lines of ISAR code but uses various lemmas. 
We now sketch how the paragraphs of the paper-and-pencil proof of
Theorem~\ref{thm:fet} translate into Isabelle.

Paragraph~\ref{M-item}: We need 22 lines of code to exhibit $M$, show its
finiteness, and exhibit $s_1$. This relies on two lemmas adding up
to Lemma~\ref{lem:set-pref-po}: 

\Snippet{liftirreflexive}
\Snippet{lifttransitive}%

The first has 7 lines of proof, but we found the second one
surprisingly hard with 160 lines of proof code. One can observe a
striking discrepancy between the shortness of the paper-and-pencil
proofs and the Isabelle proofs; probably the paper-and-pencil proofs
hide too many details, while the Isabelle proofs are more complicated
than necessary.

\DefineSnippetInline{replaceIsPreferredTruncated}{
{\isachardoublequoteopen}lessP\ les\ A\ {\isacharparenleft}replaceWithPreferred\ les\ a\ A\ U{\isacharparenright}{\isachardoublequoteclose}%
}%

\DefineSnippet{someonewinsTruncated}{\isacommand{lemma}\isamarkupfalse%
\ someone{\isacharunderscore}wins\ {\isacharcolon}\ \isanewline
\ \ \isakeyword{assumes}\ isNashWL\ {\isacharcolon}\ {\isachardoublequoteopen}isNash\ {\isacharparenleft}derivedWLGame\ gf\ Ou{\isacharparenright}\ s{\isadigit{1}}\ s{\isadigit{2}}{\isachardoublequoteclose}\ \isanewline
\ \ \isakeyword{shows}\ {\isachardoublequoteopen}{\isacharparenleft}{\isasymforall}s{\isadigit{2}}{\isacharprime}{\isachardot}\ {\isacharparenleft}form\ {\isacharquery}wlG{\isacharparenright}\ {\isacharparenleft}s{\isadigit{1}}{\isacharcomma}s{\isadigit{2}}{\isacharprime}{\isacharparenright}\ {\isacharequal}\ True{\isacharparenright}\ {\isasymor}\isanewline
\ \ \ \ \ \ \ \ \ {\isacharparenleft}{\isasymforall}s{\isadigit{1}}{\isacharprime}{\isachardot}\ {\isacharparenleft}form\ {\isacharquery}wlG{\isacharparenright}\ {\isacharparenleft}s{\isadigit{1}}{\isacharprime}{\isacharcomma}s{\isadigit{2}}{\isacharparenright}\ {\isacharequal}\ False{\isacharparenright}{\isachardoublequoteclose}%
}%

Paragraph~\ref{m-item}: we need 8 lines to exhibit $m$ and construct
$M'$. The proof of $M \prec_1^{\mathcal{P}} M'$ relies on a lemma stating 
\Snippet{replaceIsPreferredTruncated}
under the
conditions $a\in A$ and finiteness of $A$. This simple lemma has
nonetheless a proof of 17 lines.

\DefineSnippetInline{pl1notwinsTruncated}{
{\isachardoublequoteopen}{\isasymforall}s{\isadigit{1}}{\isacharprime}{\isachardot}\ {\isasymexists}s{\isadigit{2}}{\isacharprime}{\isachardot}\ {\isacharparenleft}form\ {\isacharquery}wlG{\isacharparenright}\ {\isacharparenleft}s{\isadigit{1}}{\isacharprime}{\isacharcomma}s{\isadigit{2}}{\isacharprime}{\isacharparenright}\ {\isacharequal}\ False{\isachardoublequoteclose}%
}%

Paragraph~\ref{enforce-item}: 
The fact that Player~$1$ cannot enforce $M'$ has a proof of 27 lines.
To show that Player~$2$ can therefore enforce the complement, we use 
a formalisation of one direction of Lemma \ref{lem:d-e}: 

\Snippet{determinedFormViaImplEnforceOutc}

\noindent
which has a proof of 35 lines. We then need 11 lines to prove that
Player~$2$ can actually enforce the complement. 
Another 13 lines are needed for the proof of 
$v(s_1,s_2)\in M\cap (O\backslash M')=\{m\}$. 

Paragraph~\ref{nash-item}: we show that Player~$1$ has no incentive to
deviate (32 lines) and likewise for Player~$2$ (12 lines). This implies
that we have found a Nash equilibrium.

\subsection{Remark \ref{rmk:ws-ne}}
\label{sec:ws-ne}

The right-to-left implication of Remark~\ref{rmk:ws-ne} is formalized
as follows:

\Snippet{someonewins}

\noindent
It has 35 lines of proof. The other direction,
albeit not needed for our main result, would also be interesting to
prove and we plan to do it as future work.

\section{Conclusion and future work}
\label{sec:concl}

We have formally proven an existence theorem of Nash equilibrium in
both Coq and Isabelle proof assistants. This theorem is applicable to 
every two-player game with finitely many outcomes and strict partial order preferences, provided that all derived
win/lose games from the original game are determined. Thus, the
theorem proves a transfer from winning strategies to
multi-outcome Nash equilibrium, where the latter notion is a
faithful generalization of the former.

Also, this dual formalization effort gave us the opportunity 
to sketch a comparison between Coq and Isabelle via a case study.

Regarding the underlying logic, both proof assistants rely on a
higher-order logic but that of Coq is more expressive especially
thanks to the support of dependent types, which allowed us to
formalize the basic notions of game theory in a more general way,
e.g., for an arbitrary number of players with arbitrary strategy
spaces. We suspect that generalizing the Isabelle formalization to an
arbitrary number of players would be notationally very heavy.

Regarding the proof languages, we relied on an SSReflect proof style
with a systematic use of forward-chaining in longer proofs, in order
to put forth the structure of the Coq proofs. On the other hand, the
Isabelle ISAR style allows for human-readable \emph{declarative} proof
code, but we hope to increase the degree of automation somewhat, not
too much, to avoid distraction by too much detail. Also, our
formalization work made us aware of the fact that we tend to write too
terse paper-and-pencil proofs, and suggested some improvements in the
paper-and-pencil formulation.

\medbreak
Various generalizations of the result could be proven in both proof
assistants, and some work remains to be done to benefit from the
mutual insemination between theory (paper-and-pencil proofs) and
practice (proof assistants) and between the two proof
assistants. 
We see four natural, independent directions to extend this article:
\begin{itemize}
\item Proving formally the existence of secure equilibria.

\item Since our theorem transforms determinacy into existence of NE, we plan to feed it with, e.g., the positional determinacy of parity games, which has already been formalized in Isabelle~\cite{Dittmann2016}, as mentioned in the introduction. (Technical issues may arise at the interface, though.)

\item One challenging goal would be to prove the full \cite[Theorem 2.7]{SLR2014}. As mentioned in the introduction, our result is a slight weakening of \cite[Lemma 2.4]{SLR2014}, which is the base case of the proof by transfinite induction leading to \cite[Theorem 2.7]{SLR2014}. Formalizing this proof would require us to choose a convenient representation of the countable ordinals with an associated induction proof principle. This sounds more tractable in Coq than in Isabelle.

\item Our result is weaker than \cite[Lemma 2.4]{SLR2014} for a second reason that we have not mentioned yet. This second reason is indeed orthogonal to the transfinite induction: we consider preferences that are strict partial orders instead of just acyclic binary relations. Extending our result accordingly can be done by defining a transitive closure operator, proving that it maps acyclic binary relations onto strict partial orders, and proving that an NE for bigger preferences is also an NE for smaller preferences. This sounds doable both in Coq and in Isabelle.
\end{itemize}

\bibliography{gametheo}

\end{document}